\documentclass[%
 aip,
 amsmath,amssymb,
 reprint,%
]{revtex4-1}

\usepackage{graphicx}% Include figure files
\usepackage{dcolumn}% Align table columns on decimal point
\usepackage{bm}% bold math
\usepackage[utf8]{inputenc}
\usepackage[T1]{fontenc}
\usepackage{mathptmx}

\begin{document}

\preprint{AIP/123-QED}

\title{An epistemic interpretation and foundation of quantum theory}

\author{Inge S. Helland}
 \email{ingeh@math.uio.no}
 \affiliation{Department of Mathematics, University of Oslo \\ P.O.Box 1053 Blindern, N-0316 Oslo, Norway}

%\date{\today}

\begin{abstract}

The interpretation of quantum mechanics has been discussed since this theme first was brought up by Einstein and Bohr. This article describes a proposal for a new foundation of quantum theory, partly drawing upon ideas from statistical inference theory. The approach can be said to have an intuitive basis: The quantum states of a physical system are under certain conditions in one-to-one correspondence with the following: 1) Focus on a concrete question to nature and then 2) give a definite answer to this question. This foundation implies an epistemic interpretation, depending upon the observer, but the objective world is restored when all observers agree on their observations on some variables.  The article contains a survey of parts of the author's books on epistemic processes, which give more details about the theory. At the same time, the article extends some of the discussion in the books, and at places makes it more precise. For further development of interpretation issues, I need cooperation with interested physicists.

\end{abstract}

\maketitle

\section{Introduction}

It is now a universal agreement among physicists that quantum mechanics is the most successful physical theory that has ever been developed. The calculations devised from the theory may be complicated, but there is again a universal agreement on how the calculations should be carried out. 

The history of quantum mechanics goes back to the beginning of the previous century, and the basic theory was established in the 1920s. Of fundamental importance is the treatise by von Neumann [1], which gives the basic mathematical foundation of the theory, a mathematical formulation that has been copied in dozens of textbooks.
Yet there is no agreement at all when it comes to interpretation of the theory. Many conferences on quantum foundation have been arranged in recent years, but as a result of this, the number of new interpretations has increased, and no one of the old has died out. In two of these conferences, a poll among the participants was carried out [2, 3]. The result was an astonishing disagreement on several simple and fundamental questions. One of these questions was whether quantum theory should be interpreted as an objective theory of the world (the ontic interpretation) or if it only expresses our knowledge of the world (the epistemic interpretation).

Some physicists are so frustrated about such debates that they simply say `shut up and calculate', but this is obviously not a satisfactory solution.

The main purpose of this article is to give a brief survey of the epistemic quantum interpretation contained in the author's book [4], and to discuss the implied foundational issues. A main idea behind [4] is that there should be some relationship between statistical theory and quantum theory, and that this relationship has not been properly explored earlier since there is very little communication between the corresponding scientific communities. This main idea was already formulated in the earlier book Helland [5]. The basic quantum theory ideas contained in the earlier book should only be thought about as preliminary, but on the other hand, the book included some rather deep results on group theory and its relationship to statistical inference theory.

The quantum theory of [4] in its basic assumptions has a relationship to statistical theory, and this theory will be explored  below. As an interpretation, it is related to quantum Bayesianism, see Section II, but I see subjective Bayesianism as only one possible approach to statistical theory. My interpretation is also closely related to the classical Copenhagen school. The concept of complementarity plays an important role here. The book is called `Epistemic Processes'. An epistemic process is any process where an agent intends to achieve some knowledge. It can be a statistical investigation or a quantum-mechanical measurement, but it can also take other forms.

In general, conceptual variables, variables defined by an observer or by a group of communicating observers, are important. These variables are imaged to emerge in connection to an epistemic process. Some such variables can not be measured, are inaccessible, say the full spin vector of a particle. Those which can be measured, are called epistemic conceptual variable or e-variables. In a measurement setting they are thought as analoguous to the parameters of statistical models; see Section III. As will be discussed below, the states of the physical system have interpretations related to a question `What is the value of $\theta$?' for an e-variable $\theta$, together with some information about $\theta$, in the simplest case full information $\theta=u$.

Recently, Chiribella et al. [6] addressed the question whether or not quantum theory could be reconstructed from the Bayesian interpretation. A citation from that paper may be of interest: `Each interpretation comes with its own narrative, and, if the narrative is powerful enough, the mathematical framework of quantum theory should emerge directly from it, without any further assumption. Unfortunately, the current status of the quantum interpretation is strikingly different: the interpretations are inspired by the mathematical framework, but none of them has been shown to be sufficient to reconstruct it.'

The problem of reconstructing the Hilbert space formalism from the epistemic interpretation will be briefly discussed in Section IV below; a more thorough discussion is given in Helland [7].

\section{Relation to qBism and to Niels Bohr's complementarity concept}

The interpretation discussed in [4] is related to quantum Bayesianism or qBism [8-10], but it differs at some points. The main common thought is that the observer(s) and his (their) mind(s) play(s) an important role. The predictions of quantum mechanics involve probabilities, and a QBist interpret these as purely subjective probabilities, attached to a concrete agent, or observer. There are many elements of QBism which represent something completely new, both in relation to classical physical theory, in relation to many people's conceptions of science in general, and also in relation to earlier interpretations of quantum mechanics. The essential assumption is that the observer plays a role which cannot be eliminated. 

Subjective Bayes-probabilities have also been in fashion among groups of statisticians. My strong opinion is that it can be fruitful to look for analogies between statistical inference theory and quantum mechanics, but then one must look more broadly upon statistics and statistical inference theory, not only focus on subjective Bayesianism. This is only one of several philosophies that can form a basis for statistics as a science. There are also different versions of Bayesianism. The present paper is closest to the use of Bayes' theorem with a prior based on symmetry. The ideal observer may well be a Bayesian, but for real observers other views upon statistical inference will also be relevant. Statistical inference is concerned with data and models for these data. To reach conclusions it is essential to have a theory of decisions.

In [4] `decision' is taken as a primitive concept. A decision can either be taken as resulting from a (subjective) belief as in qBism, but it also can be a result of firm knowledge (or it may be a combination). The process behind obtaining knowledge can take very many different forms.

In my opinion, this view is also valid for quantum theory. One of my points of departure is that I look upon a quantum state as the result of two decisions: A decision to focus upon a question to nature, and a decision to interpret and express the answer. 

In discussing these and similar questions, it can be useful to have a closer look at Niels Bohr's concept \emph{complementarity}. For a thorough discussion of complementarity in physics, see  Plotnitsky [11]. The concept was originally introduced by Niels Bohr to describe what it is possible to measure physically, but in various talks Bohr also looked upon extensions of the complementarity concept. Such extensions are also of great current interest.

Here is Plotnitsky’s definition of complementarity:

(a)	a mutual exclusivity of certain phenomena, entities, or conceptions; and yet

(b)	 the possibility of applying each one of them separately at any given point; and

(c)	the necessity of using all of them at different moments for a comprehensive account of the totality of phenomena that we consider.

Many physics papers discuss two actors Alice and Bob, being so far away from each other that they do not communicate. All physicists agree that there exist situations where the observations of Alice and Bob are entangled. From an epistemic point of view the two actors may also have complementary comprehensions of the world because they focus differently. According to John A. Wheeler, each observer can create his/her own history.

I claim that this may be equally true for persons – or groups of persons – making experiences in the macroworld. People may tend to have different world views. Examples of this may be given, and are given in [4], but this is beyond the scope of the present article. 

\section{Interpretation of the quantum state}

\subsection{Conceptual variables} 

A conceptual variable is any variable defined by an observer or by a group of communicating observers. I will assume that each conceptual variable varies on some topological space. For a physical system under measurement, two kinds of conceptual variables exist. 

A variable may be accessible, possible to measure, like a velocity or a spin-component of a particle in some given direction. Such variables are called epistemic conceptual variables, \emph{e-variables}, and are in my view closely related to the parameters of statistical theory. 

Or they may be inaccessible, like the full spin-vector or the vector (position, velocity). When a vector $\phi=(\theta^1,\theta^2 )$ is inaccessible, but the components $\theta^1$ and $\theta^2$ are e-variables, we are in a situation where we have a choice of measurement, and we might say that $\theta^1$ and $\theta^2$ are complementary.

\subsection{State vectors corresponding to maximally accessible e-variables}

Assume a finite-dimensional Hilbert space $H$, and let $|v\rangle$ be some unit vector in $H$. Then trivially, $|v\rangle$ is an eigenvector of many operators. Assume that one can find such an operator $A$ satisfying 1) $A$ is physically meaningful, that is, can be associated with an e-variable $\theta$; 2) $A$ has only one-dimensional eigenspaces. 

Then in particular, $|v\rangle$ corresponds to a single eigenvalue $u$ of $A$, and can be associated with a question: `What is the value of $\theta$?' together with an answer: $\theta=u$.

It is easy to see, and is shown explicitly in Helland [7], that all eigenspaces of $A$ are one-dimensional if and only if $\theta$ is maximally accessible: Whenever $\theta=f(\zeta)$ for a function $f$ which is not one-to-one, the conceptual variable $\zeta$ is inaccessible.

\subsection{General state vectors}

If the situation is as in Subsection 3.2, but $A$ is more general, then the \emph{eigenspaces} of $A$ can be associated with a question-and-answer pair concerning $\theta$.

\subsection{Spin and angular momentum}

The interpretation given above was abstract and vague. For the case of spin/angular momentum, more concrete interpretations can be given, at least for certain state vectors. To see this, consider a spin or angular momentum vector $\phi$ with fixed norm varying on a sphere $\Phi$. This vector is inaccessible. However, given some direction $a$, the components $\theta^a =a\cdot \phi$ can be measured and are e-variables. Given a certain normalization of $\phi$, each $\theta^a$ takes the values $-j, -j+1,...,j-1, j$ for some integer or half-integer $j$.
\bigskip

\textbf{Proposition 1} [4]

\textit{Assume the usual Hilbert space for spin/angular momentum. For each $a$ and each $k$ (k=-j,...,j) there is exactly one normalized ket vector $|v\rangle=|a;k\rangle$ with arbitrary phase such that the operator $J^a$ corresponding to $\theta^a$ satisfies $J^a |v\rangle=k|v\rangle$. This ket vector corresponds to the question `What is the value of the angular momentum component $\theta^a$?' together with the definite answer `$\theta^a =k$'.}
\smallskip

For the qubit case, dimension 2 of the Hilbert space, these vectors constitute all ket vectors. This can be seen by a simple Bloch sphere argument.

\subsection{A tentative general theorem}

Let in general $\phi$ be an inaccessible conceptual variable taking values in some topological space $\Phi$, and let $\theta^a =\theta^a (\phi)$ be accessible functions for $a$ belonging to some index set $\mathcal{A}$. Assume that each $\theta^a$ is maximally accessible, and assume that there is a one-to-one relationship between the different e-variables: For $a\ne b$ there exists an invertible transformation $k_{ab}$ such that $\theta^b (\phi)=\theta^a (k_{ab}\phi)$ (no summation convention). The spin/angular momentum situation is a special case of this. In general, $\theta^a$ varies over a space $\Theta^a$,

For each $a$, let $G^a$ be the group of automorphisms on $\Theta^a$, and for $g^a \in G^a$ let $k^a$ be any transformation on $\Phi$ for which $g^a \theta^a(\phi)=\theta^a (k^a \phi)$. It is easy to verify that for fixed $a$ the transformations $k^a$ form a group $K^a$. 

Let $K$ be the group on $\Phi$ generated by the $K^a$'s and the elements $k_{ab}$.

Make the following assumptions:

1) The group $K$ is a locally compact topological group satisfying weak assumptions such that an invariant measure $\mu$ on $\Phi$ exists.

2) The group generated by products of elements in $K^a ,K^b,....; a,b,...\in\mathcal{A}$ is equal to $K$.

Consider the case where each $\theta^a$ takes a finite $d$ number of values $\{u_k\}$.

In the spin/angular momentum case, all these assumptions are satisfied. The toy model of Spekkens [12] satisfies the assumptions except 2) above.

Now fix one index $0\in\mathcal{A}$ and consider the Hilbert space

\begin{equation}
H=\{f\in \mathrm{L^2}(\Phi,\mu ): f(\phi)=\tilde{f}(\theta^0 (\phi))\}.
\label{H}
\end{equation}
This Hilbert space is $d$-dimensional.
\bigskip

\textbf{Theorem 1} [4]

\textit{Under some extra technical conditions the following holds: For every $a, u_k$ and associated with every indicator function $I(\theta^a (\phi)=u_k )$ there is a vector $|a;k\rangle \in H$. The mapping $I(\theta^a (\phi)=u_k )\mapsto |a;k\rangle$ is invertible in the sense that $|a;k\rangle \ne |b;j\rangle$ for all $a,b,k,j$ except in the trivial case $a=b; j=k$.  This inequality is interpreted to mean that there is no phase factor $e^{i\gamma}$ such that $|a;k\rangle =e^{i\gamma}|b;j\rangle$. For each $a$ the vectors $|a;k\rangle$ form an orthonormal basis of $H$.}
\smallskip

This means that the vectors $|a;k\rangle$ can be interpreted as a question: `What is the value of $\theta^a$?' together with the answer: `$\theta^a =u_k$'.

It is argued in [4] that the technical conditions given there are not the best possible. It is an open question how they can be weakened.

\subsection{Preparations and ideal measurements}

In the simplest case of measurement the physical system is first prepared in a state $|a;k\rangle$, which thus under certain conditions can be interpreted as associated with a question-and-answer pair. More generally, the system can be prepared in an eigenspace $V_{ka}$ of a corresponding operator $A^a$, with a similar interpretation. Still more generally, the observer only knows a probability distribution $p_{ka}$ over the different values of the e-variable $\theta^a$. This is equivalent to specifying a density operator

\begin{equation}
\rho^a =\sum_k p_{ka} \Gamma_{ka},
\label{rho}
\end{equation}
where $\Gamma_{ka}$ is the projector upon $V_{ka}$.

An ideal measurement can now be described by posing a new question `What is the value of $\theta^b$?' for another e-variable $\theta^b$, and at the end obtaining a definite answer $\theta^b =v_j$. Formally, this can be connected to a resolution of the identity

\begin{equation}
I=\sum_j \Gamma_{jb}
\label{identity}
\end{equation}
and an operator for $\theta^b$

\begin{equation}
A^b=\sum_j v_j \Gamma_{jb}.
\label{operator}
\end{equation}

The $v_j$'s are assumed to be distinct values. Equivalently to ask for values of $\theta^b$, one can ask for values of $\psi^b =f(\theta^b)$ for a one-to-one function $f$.

\subsection{Real measurements}

Every measurement needs a measurement apparatus, and such an apparatus is usually not perfect. A statistician would distinguish between the data (the result of the measurement) on the one hand, and the ideal measurement on the other hand, which is taken as a parameter in the statistical measurement model. In quantum theory this distinction is not always clear. The e-variables of elementary quantum theory are  discrete, and experiments are often very accurate. From a statistical point of view this results in confidence intervals, credibility intervals or prediction intervals which in general may be taken to include the true value, but which in this case degenerate into a single point, which then necessarily must  be equal to the true value. Thus in such cases the distinction between estimates and true value is blurred out, and much of statistical theory becomes irrelevant.

In this subsection we will look at the situation where we do have to take measurement error into account. An important point is that the e-variables can be thought of as analoguous to the parameters of statistics. I could have used the notion `parameter' throughout, but this word is unfortunately overburdened in physics, and has usually another meaning there.

  A statistician would device a statistical model for the measurements performed by the measurement apparatus. This is a probability model for the data, given the parameters. Assuming discrete e-variables/parameters and discrete data, this is a point probability $q(x|\theta^b)$ for the data $x$. Given the resolution of the identity (\ref{identity}) we can first define a likelihood effect by

\begin{equation}
L=L(x)=\sum_j q(x|\theta^b=v_j ) \Gamma_{jb}.
\label{effect}
\end{equation}

In [4] is proved a focused likelihood principle, stating that the question asked and the experimental evidence on the answer are functions of $L$.

A positive operator-valued measure on the data space is given by $M(C)=\sum_{x\in C} L(x)$. By a slight generalization of Born's formula the probability distribution of the data from the experiment sketched in Subsection F is given by

\begin{equation}
P(x\in C|\rho^a )=\mathrm{trace}(\rho^a M(C)).
\label{inference}
\end{equation}
This can be taken as a starting point for quantum inference.

\section{On the reconstruction of quantum theory from the epistemic interpretation}

In several recent investigations, for instance [13] and [14], quantum theory has been reconstructed from sets of plausible axioms. However, as indicated in [6], the link between these axioms and different interpretations seems to lacking to some extent.

In Helland [7] the Hilbert space structure of quantum theory was deduced from a setting with conceptual variables, where focusing and symmetry were basic assumptions. First, it is assumed a space $\Phi$ upon which an inaccessible variable $\phi$ varies, and a group $K$ acting upon this space. Next, assume an accessible conceptual variable, an e-variable, $\theta=\theta(\phi)$. This function is assumed to be \emph{permissible}:

\begin{equation}
\theta(\phi_1)=\theta(\phi_2)\ \mathrm{implies}\ \theta(k\phi_1)=\theta(k\phi_2)\ \mathrm{for\ all}\ k\in K.
\label{permissible}
\end{equation}
This implies that a group $G$ on the range $\Theta$ of $\theta$ may be defined by

\begin{equation}
(g\theta)(\phi):=\theta(k\phi); k\in K.
\label{G}
\end{equation}
The function from $K$ to $G$ defined by (\ref{G}) is a group homomorphism.

Assume that $K$ is transitive on $\Phi$, and fix $\phi_0 \in \Phi$. Then every $\phi \in \Phi$ can be written as $\phi =k\phi_0$ for some $k\in K$.

Next turn to the theory of coherent states; see [15]. Under weak assumptions we can find a Hilbert space $H$ and an irreducible representation $V$ of $K$ on this Hilbert space such that the following holds: Fixing $|\phi_0\rangle\in H$, and defining $|\phi\rangle=|\phi(k)\rangle=V(k)|\phi_0\rangle$ when $\phi=k\phi_0$ gives a square-integrable coherent state system. 

Weak assumptions are made such that there is a left-invariant measure $\nu$ on $\Phi$, and then by using Schur's Lemma there is a resolution of the identity

\begin{equation}
\int |\phi\rangle\langle \phi | d\mu (\phi) =I
\label{identity1}
\end{equation}
for a measure $\mu$ satisfying $d\mu (\phi)=\lambda^{-1}d\nu (\phi)$ for some $\lambda > 0$.

The group $G$ on $\Theta$ is not necessarily transitive, but now I use the following recent principle from statistics [4]: Every model reduction of a statistical model should be to an orbit or to a set of orbits of the group when a group is defined on the parameter space.

For the e-variable $\theta$ I assume reduction to a single orbit of $G$, so that $G$ now is transitive. Then fix $\theta_0\in\Theta$ (the reduced space), and repeat the theory above. This gives a resolution of the identity

\begin{equation}
\int |\theta\rangle\langle \theta | d\rho (\theta)
\label{identity2}
\end{equation}
for a measure $\rho$. The operator $A$ corresponding to $\theta$ can be defined as

\begin{equation}
A=\int \theta |\theta\rangle\langle \theta | d\rho (\theta)
\label{A}
\end{equation}
with a suitable domain of definition.

For the case when $A$ has a discrete spectrum, it is proved in [7] that the possible values of $\theta$ coincide with the eigenvalues of $A$.

Of course, reconstructing the Hilbert space structure is only part of reconstructing the full quantum theory. In [4] the Born formula is proved from two assumptions: 1) The focused likelihood principle mentioned above; 2) An assumption of rationality as formulated by the Dutch Book principle. There are also several other derivations of Born's formula from reasonable assumptions; see for instance [14]. The Schr\"{o}dinger equation is derived in the one-dimensional case by considering the particle trajectory as an inaccessible conceptual variable, and conditioning on the past and on the future in this trajectory. It seems plausible that further reasonable assumptions will give the full theory as formulated for instance by Volovich [16].

\section{The epistemic interpretation: Resolving paradoxes}

The present article and [4] give an essentially new approach to quantum mechanics: A quantum
state is associated with a focused question to nature and an answer to that question. The observer plays a role in determining this state, but nevertheless an
objective world emerges when all real and imagined observers agree. We will see how this simplifies
interpretation in three examples usually associated in different ways with the phrase `quantum
paradoxes'.
\bigskip

\underline{Example 1. Schr\"{o}dinger's cat.} The discussion of this example concerns the state of the cat just
before the sealed box is opened. Is it half dead and half alive?

To an observer outside the box the answer is simply: He doesn't know. Any accessible e-variable
connected to this observer does not contain any information about the death status of the cat. But on
the other hand – an imagined observer inside the box, wearing a gas mask, will of course know the
answer. The interpretation of quantum mechanics is epistemic, not ontic, and it is connected to the
observer. Both observers agree on the death status of the cat once the box is opened.
\bigskip

\underline{Example 2. Wigner’s friend.} Was the state of the system only determined when Wigner learned the
result of the experiment, or was it determined at some previous point?

My answer to this is that at each point of time there is one quantum state connected to Wigner’s
friend as an observer and one quantum state connected to Wigner, depending on the knowledge that
they have at that time. The superposition given by formal quantum mechanics corresponds to a `don't
know' epistemic state. The states of the two observers agree once Wigner learns the result of the
experiment.
\bigskip

\underline{Example 3. The two-slit experiment.} This is an experiment where all real and imagined observers
communicate at each point of time, so there is always an objective state. 

Look first at the situation when we do not know which slit the particle goes through. This is really
a `don't know' situation. Any statement to the effect that the particles somehow pass through both
slits, see e.g. [17], is meaningless. The interference pattern can be explained by the fact that the particles are (nearly)
in an eigenstate in the component of momentum in the direction perpendicular to the slits in the plane
of the slits. If any observer finds out which slit the particles goes through, the state changes into an
eigenstate for position in that direction. In either case the state is an epistemic state for each of the
communicating observers, and thus also an ontic state.

\section{Concluding remarks}

The notion of conceptual variables also has links to other interpretations of quantum theory. Take for instance the classical Bohm interpretation, constructed from a particle trajectory plus a pilot wave. These constructions are just conceptual variables, but at least the full trajectory must be inaccessible. Or take the many worlds/ many minds interpretations: Here the different worlds must be considered as conceptual variables, but only one world is accessible.

The present treatise is based on focusing, symmetry and decisions. These concepts are not confined to the micro-world. This is consistent with the fact that quantum theory recently has been applied in cognitive models and in certain economic models; see references in [4].

I do not regard the present paper as the final solution to the quantum interpretation problem, but I do think it goes a long way towards this solution. For further work towards the final solution, I need cooperation with interested physicists. I define myself as a statistician. A general point is that cooperation between different scientific disciplines is becoming increasingly important in our modern time.

 \section*{Acknowledgment}
 
 I am grateful to Bj\o rn Solheim for general discussions and for providing the references [6] and [14].

\section*{References}

\setlength\parindent{0cm}

[1] J. von Neumann, \textit{ Mathematische Grundlagen der Quantenmechanik.} (Springer, Berlin, 1932)

[2] M. Schlossbauer, J. Koller  and A. Zellinger,  A snapshot of fundamental attitudes toward quantum 
         mechanics. Studies in History and Philosophy of Science \textbf{44}(3), 222-230 (2013)
         
[3] T. Norsen and S. Nelson, Yet another snapshot of fundamental attitudes toward quantum 
          mechanics. arXiv: 1306.4646 [quant-ph] (2013)

[4] I.S. Helland,  \textit{Epistemic Processes. A Basis for Statistics and Quantum Theory.} (Springer, Berlin, 2018)

[5] I.S. Helland, \textit{Steps Towards a Unified Basis for Scientific Models and Methods.} (World Scientific, Singapore, 2010).

[6] G. Chiribella, A. Cabello, M. Kleinmann and M.P. M\"{u}ller, General Bayesian theories and the emergence of the exclusivity principle. arXiv: 1901.11412v2 [quant-ph] (2019)

[7] I.S. Helland, Symmetry in a space of conceptual variables. J. Math. Phys. 60 (5) (2019)

[8] C.M. Caves, C.A. Fuchs and R. Schack, Quantum probabilities as Bayesian probabilities. Phys. Rev. A \textbf{65}. 022305 (2002)

[9] C.A. Fuchs, QBism, the perimeter of quantum Bayesianism. arXiv: 1003.5209 [quant-ph] (2010)

[10] C.A. Fuchs and R. Schack, Quantum-Bayesian consistency. Rev. Mod. Phys. \textbf{85}, 1603 (2013)

[11] A. Plotnitsky, \textit{Niels Bohr and Complementarity. An Introduction.} (Springer, New York, 2013)

[12] R.W. Spekkens, In defense of the epistemic view of quantum states; a toy theory. Phys. Rev. A \textbf{75}, 032110 (2007)

[13] L. Hardy, Quantum theory from five reasonable axioms. arXiv: 0101012v4 [quant-ph] (2001)

[14] G. Chiribella, G.M. D'Ariano and P. Perinotti, Quantum from principles. In: \textit{Quantum Theory: Informational Foundations and Foils} G. Chiribella and P.W. Spekkens [Ed.] pp. 171-221 (Springer, Berlin, 2016)

[15] A. Perelomov, \textit{Generalized Coherent States and Their Applications.} (Springer, Berlin, 1986)

[16] I.V. Volovich, Seven principles of quantum mechanics. arXiv: 0212126v1 [quant-ph] (2002)

[17] A. Ananthaswamy, \textit{Through Two Doors at Once.} (Dutton, New York, 2018)

\end{document}